\documentclass[%
 reprint,
 amsmath,amssymb,
 aps,
]{revtex4-1}

\usepackage{graphicx}
\usepackage{dcolumn}
\usepackage{bm}

\newcommand{\bea}{\begin{eqnarray}}	
\newcommand{\eea}{\end{eqnarray}}
\newcommand{\be}{\begin{equation}}	
\newcommand{\ee}{\end{equation}}
\newcommand{\beq}{\begin{equation}}	
\newcommand{\eeq}{\end{equation}}

\newcommand{\Z}{{\mathbb Z}}
\newcommand{\C}{{\mathbb C}}

\newcommand{\vev}[1]{\left\langle{#1}\right\rangle}

\def\R{\relax\ifmmode {\mathbb R}  \else${\mathbb R}$\fi}
\def\C{\relax\ifmmode {\mathbb C}  \else${\mathbb C}$\fi}
\def\Z{\relax\ifmmode {\mathbb Z}  \else${\mathbb Z}$\fi}
\def\N{\relax\ifmmode {\mathbb N}  \else${\mathbb N}$\fi}
\def\I{\relax\ifmmode {\mathbb I}  \else${\mathbb I}$\fi}

\begin{document}


\title{A first survey of the ghost-gluon vertex in the Gribov-Zwanziger framework }
\author{Bruno W. Mintz}
 \altaffiliation{Invited talk presented at XIV International Workshop on Hadron Physics, Florian\'opolis, Brazil, March 2018.}
\author{Leticia F. Palhares}%
\author{Silvio P. Sorella}
\affiliation{%
 UERJ - Universidade do Estado do Rio de Janeiro\\
 Departamento de Fisica Teorica,\\
 Rua Sao Francisco Xavier, 524\\
 20550-900, Maracana, Rio de Janeiro, Brasil
}%


\author{Antonio D. Pereira}
\affiliation{
 Institut f\"ur Theoretische Physik, Universit\"at Heidelberg\\
 Philosophenweg 12, 69120 Heidelberg, Germany
}%

\date{\today}

\begin{abstract}

As the restriction of the gauge fields to the Gribov region is taken into account, it turns 
out that the resulting gauge field propagators display a nontrivial infrared behavior, being 
very close to the ones observed in lattice gauge field theory simulations. In this work, we 
explore for the first time a higher correlation function in the presence of the Gribov horizon: 
the ghost-anti-ghost-gluon interaction vertex, at one-loop level. Our analytical results (within 
the so-called Refined Gribov Zwanziger theory) are fairly compatible with lattice YM simulations, 
as well as with solutions from the Schwinger-Dyson equations. This is an indication that the RGZ
framework can provide a reasonable description in the infrared not only of gauge field propagators, 
but also of higher correlation functions, such as interaction vertices.


\end{abstract}

\maketitle


\section{\label{sec:introduction}Introduction}

In order to have a proper definition of a quantum nonabelian gauge theory, such as QCD, one 
needs to have a well-defined generating functional. In the perturbative regime at high energies, 
this can be achieved through the well-known Faddeev-Popov quantization procedure
\cite{Faddeev:1967fc}. For energy 
scales in which the coupling of the strong force becomes sufficiently high, the perturbative 
approach may display some inconsistencies. One such problem is the so-called Gribov problem: 
in a gauge fixed theory, the ghost fields acquire a pole at finite (euclidean) momentum. 
Equivalently, the determinant of the Faddeev-Popov operator vanishes and the generating 
functional of the theory becomes void. In order to correct this problem (at least partially), 
Gribov suggested to restrict the functional integration of gauge fields to a subset of the 
field space, which is now called the Gribov region \cite{Gribov:1977wm}.

The restriction of the gauge fields to the Gribov region can be effectively implemented 
by the so-called Horizon function, introduced by Zwanziger in \cite{Zwanziger:1989mf}, 
which is added to the Yang-Mills action according to the Lagrange multiplier method. 
However, as the fields are constrained to the Gribov region, the theory develops nonzero 
expectation values of dimension two operators, i.e., the condensates $\vev{A_\mu^aA_\mu^a}$
and $\vev{\bar\varphi_\mu^{ab}\varphi_\mu^{ab}-\bar\omega_\mu^{ab}\omega_\mu^{ab}}$ 
are nonzero in the presence of the 
Gribov horizon. This gives rise to the so-called Refined Gribov-Zwanziger (RGZ) action 
\cite{Dudal:2008sp,Dudal:2011gd}, whose tree-level propagator displays a close 
agreement with Monte Carlo lattice simulations \cite{Dudal:2010tf,Cucchieri:2011ig}
and other nonperturbative methods, as the Schwinger-Dyson equations 
\cite{Aguilar:2013xqa,Aguilar:2015bud,Bashir:2012fs} and the Functional 
Renormalization Group \cite{Berges:2000ew,Pawlowski:2005xe}.

Given the success of the RGZ framework in the description of the YM propagators
already at tree-level, it is natural to ask if higher correlation functions 
(i.e., vertices), will also display a reasonable behavior, especially at the 
low momentum region, where nonperturbative features are expected to be more relevant.

In this work, we calculate the ghost-gluon vertex at the one-loop approximation 
in the RGZ framework at the soft-gluon kinematic regime. In Section \ref{sect:RGZ},
we briefly review the RGZ action, while in Section \ref{sect:vertex} we discuss 
some details of the calculation of the ghost-gluon vertex. Next, we show and discuss 
our results in Section \ref{sect:results} and finally end with some remarks 
in Section \ref{sect:remarks}.

\section{The RGZ framework}\label{sect:RGZ}

The starting point of the construction of the RGZ action is, of course, the 
Yang-Mills action \cite{Peskin:1995ev,Weinberg:1996kr,Yang:1954ek} 
quantized in some 
fixed gauge according to the Faddeev-Popov prescription. The theory is defined
in an euclidean flat space\footnote{The continuation to Minkowski space of 
the correlation functions of the theory is generally rather delicate due to 
the appearance of singularities (poles and cuts) in the complex plane.}, 
so that one does not distinguish between covariant 
and contrariant vectors. In the Landau gauge, the Faddeev-Popov action reads
\begin{eqnarray}\label{eq:FP-action}
 S_{FP} = \int d^dx\,\left[\frac14F_{\mu\nu}^aF_{\mu\nu}^a + \bar c^a\partial_\mu\,D_\mu^{ab}c^b
 +ib^a\partial_\mu A_\mu^a\right],
\end{eqnarray}
where
\begin{eqnarray}
 F_{\mu\nu}^a = \partial_\mu A_\nu^a - \partial_\nu A_\mu^a + gf^{abc}A_\mu^bA_\nu^c
\end{eqnarray}
is the gluon field strength,
\begin{eqnarray}
 D_\mu^{ab} = \delta^{ab}\partial_\mu - gf^{abc}A_\mu^c
\end{eqnarray}
is the covariant derivative in the adjoint representation of the color group 
$SU(N_c)$ and $b$ is the Nakanishi-Lautrup field that implements the Landau 
gauge condition $\partial_\mu A^a_\mu=0$. The color indices run from zero up
to $N_c^2-1$, since all fields are in the adjoint representation of $SU(N_c)$.

Following Zwanziger \cite{Zwanziger:1989mf,Vandersickel:2012tz,Vandersickel:2011zc}, 
the restriction to the Gribov 
horizon can be implemented at the level of the 
generating functional by requiring that the expectation value of the 
so-called Zwanziger's horizon function (related to the trace 
of the Faddeev-Popov operator) is constrained to be
\begin{eqnarray}\label{eq:horizon-function}
 h(A) \equiv \int\,d^dx\, gf^{abc}A_\mu^b\,[{\cal M}^{-1}(A)]^{cd}gf^{aed}A_\mu^e = dV(N_c^2-1),\nonumber\\
\end{eqnarray}
where $d$ is the dimension of (euclidean) spacetime, $V$ is its volume, and
\begin{eqnarray}
 [{\cal M}(A)]^{ab} = -\partial_\mu\,D_\mu^{ab}
\end{eqnarray}
is the Faddeev-Popov operator.  In order to impose the horizon constraint 
(\ref{eq:horizon-function}) at the level of the path integral, one should 
use the Lagrange multiplier method, so that the constraint is implemented 
by adding 
\begin{eqnarray}
 S_h = \gamma^4\left[h(A) - dV(N_c^2-1)\right]
\end{eqnarray}
to the Faddeev-Popov action (\ref{eq:FP-action}) to get the Gribov-Zwanziger 
(GZ) action
\begin{eqnarray}
 S_{GZ} = S_{FP} + S_h.
\end{eqnarray}

Note that the horizon function (\ref{eq:horizon-function}) contains the inverse 
of a differential operator (i.e., {\cal M}). Therefore, it corresponds to a nonlocal 
term in the action. However, the horizon function can be rewritten within the 
generating functional in a local manner, with the introduction of auxiliary fields, 
the bosonic 
fields $\bar\varphi_\mu^{ab}$ and $\varphi_\mu^{ab}$, and the fermionic fields 
$\bar\omega_\mu^{ab}$ and $\omega_\mu^{ab}$. In terms of these auxiliary fields, 
an equivalent local horizon function emerges, which is given by
\begin{eqnarray}\label{eq:local-horizon}
 S_H &=& \int\,d^dx\,\left[\bar\varphi_\mu^{ab}\,[{\cal M}(A)]^{bc}\varphi_\mu^{cb}
 - \bar\omega_\mu^{ab}\,[{\cal M}(A)]^{bc}\omega_\mu^{cb} + \right.\nonumber\\
 &&\left.+ g\gamma^2f^{abc}A_\mu^a(\bar\varphi_\mu^{bc}-\varphi_\mu^{bc})\right],
\end{eqnarray}
so that the local GZ action is given by $S_{GZ}=S_{FP}+S_H$. 

Even though the GZ action does constrain the fields to within the Gribov horizon, 
it is not complete, in the sense that it is unstable with respect to the 
formation of the condensates $\langle A_\mu^aA_\mu^a\rangle$ and 
$\langle\bar\varphi_\mu^{ab}\varphi_\mu^{ab}-\bar\omega_\mu^{ab}\omega_\mu^{ab}\rangle$,
which are energetically favored, as shown in the analyses of 
\cite{Dudal:2008sp,Dudal:2011gd}. Thus, using once again the Lagrange multiplier 
method, one can introduce extra terms in the action in order to account for the 
nonvanishing condensates we just mentioned. The extra terms are
\begin{eqnarray}
 S_{cond} = \int\,d^dx\,\left[\frac{m^2}{2}A_\mu^aA_\mu^a
 -M^2(\bar\varphi_\mu^{ab}\varphi_\mu^{ab}-\bar\omega_\mu^{ab}\omega_\mu^{ab})\right].
\end{eqnarray}

Finally, with the condensates properly accounted for, one arrives at the Refined 
Gribov-Zwanziger action 
\begin{eqnarray}\label{eq:RGZ-action}
 S_{RGZ} = S_{FP} + S_{H} + S_{cond},
\end{eqnarray}
which will be used throughout this work. With this action, the tree level gluon propagator 
is modified with respect to perturbative Yang-Mills, since it has an effective mass term
(coming from the condensate $\langle A^2\rangle$) and it also couples linearly with 
the auxiliary fields $\bar\varphi$ and $\varphi$. As a result, the tree-level RGZ 
momentum space gluon propagator (in the Landau gauge) reads
\begin{eqnarray}\label{eq:gluon-propagator}
 D_{\mu\nu}^{ab}(p) = \delta^{ab}\left(\delta_{\mu\nu}-\frac{p_\mu p_\nu}{p^2} \right)D(p^2),
\end{eqnarray}
where the tree-level RGZ gluon form factor is given by
\begin{eqnarray}
 D(p^2) = \frac{p^2+M^2}{p^4+(m^2+M^2)p^2 + m^2M^2+2g^2N_c\gamma^4}.
\end{eqnarray}
As mentioned in the introduction, the gluon propagator (\ref{eq:gluon-propagator}) 
calculated with the RGZ action displays an excellent agreement 
with nonperturbative methods in gauge theory. This is a motivation for taking the RGZ 
action seriously as an effective model for the strong interactions. In the next section, 
we shall calculate a higher correlation function, that is, the ghost-gluon vertex at 
one-loop level from a perturbative expansion of the RGZ action (\ref{eq:RGZ-action}).

\section{The one-loop vertex function}\label{sect:vertex}

Starting from the RGZ action (\ref{eq:RGZ-action}), one can use standard QFT techniques 
to calculate the Feynman rules (i.e., propagators and bare vertices) of the theory.
From these, the calculation of correlation functions can be made systematically 
using the language of Feynman diagrams. We are particularly interested in the 
ghost-gluon vertex, given by the connected function
 \begin{eqnarray}\label{eq:cbcA-def}
 \vev{A_\mu^a(k)\,\bar c^b(p)\,c^c(q)}_{q=-p-k} = \left.\frac{\delta^3Z_c}{\delta (J_{ A})_\mu^a(k)\delta J_{\bar c}^b(p)\delta J_{c}^c(q)}\right|_{q=-p-k},
\end{eqnarray}
where $Z_c$ is the generating functional of the connected correlation functions, and 
$J_A$, $J_{\bar c}$ and $J_c$ are, respectively, the sources for the gauge field $A$, 
for the ghost field $c$, and for the antighost field $\bar c$, which are all taken 
to zero after the functional derivatives are taken. 

\begin{figure}
\includegraphics[height=6.5cm]{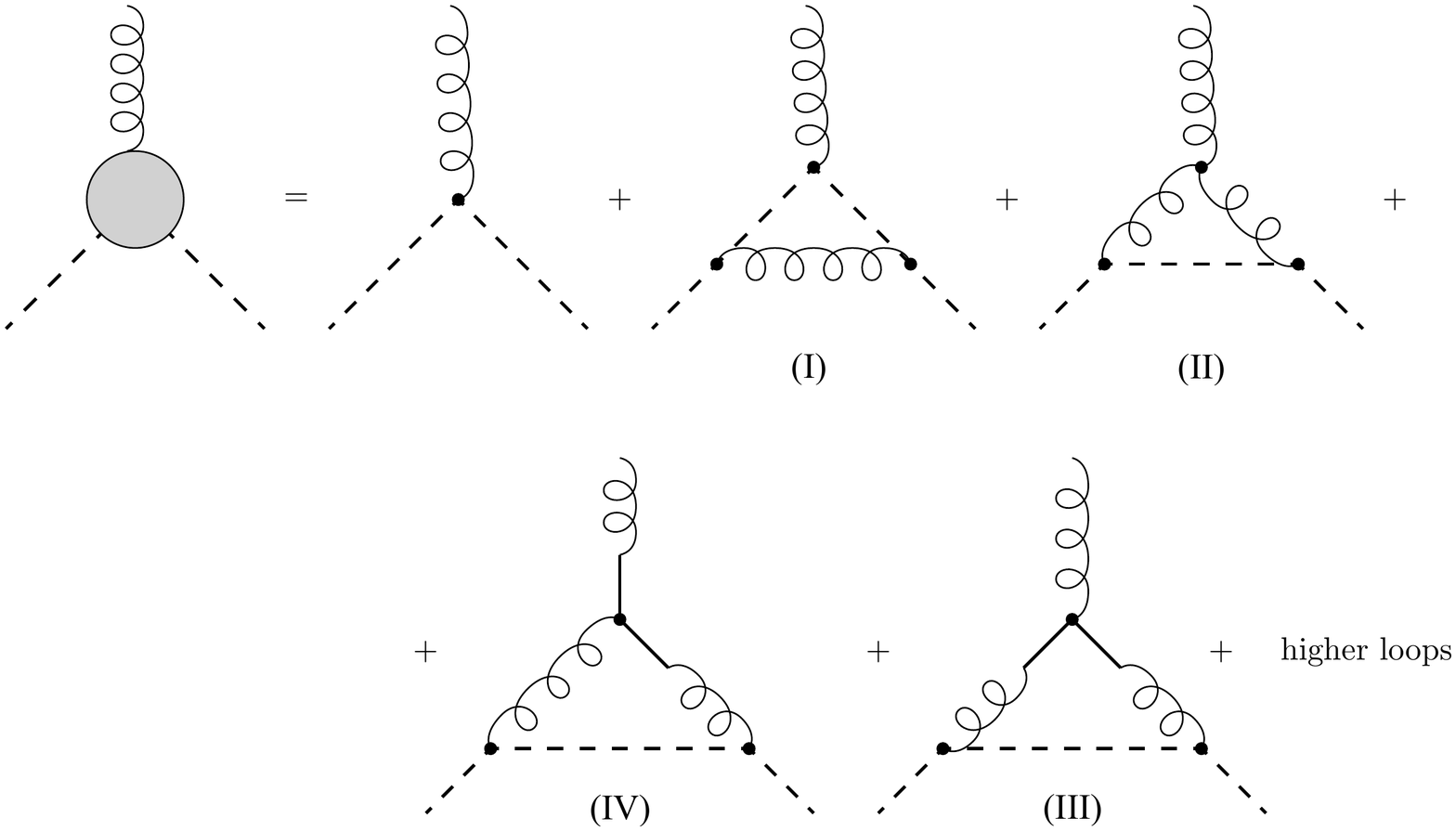}
\vspace{-.8cm}
\caption{Feynman diagram expansion up to one-loop order for the ghost-gluon vertex in the Refined Gribov-Zwanziger theory. Dashed lines represent ghosts and antighosts, while the curly lines stand for gluons. Full lines,  that only appear in mixed propagators, correspond to the auxiliary fields $\varphi,\bar\varphi$.}
\label{fig:feyndiags}
\end{figure}

Diagramatically, the vertex (\ref{eq:cbcA-def}) at one-loop can be represented by 
the Feynman diagrams of Fig. \ref{fig:feyndiags}. Note that, besides the first three
diagrams, which are also present in the perturbative Yang-Mills calculation
\cite{Ball:1980ax}, the restriction to the Gribov horizon give rise to two extra 
diagrams (III and IV), in which the auxiliary fields $\bar\varphi$ and $\varphi$ 
appear in mixed propagators $\vev{A\bar\varphi}$ and $\vev{A\varphi}$ with 
the gluon field and also in the $A-\bar\varphi-\varphi$ 
vertex. These extra diagrams can be interpreted as the effect of the horizon 
function. If instead of using the local formulation of the GZ action 
(\ref{eq:local-horizon}), one insisted in using the nonlocal horizon function 
(\ref{eq:horizon-function}), the vertices with the auxiliary fields would correspond
to vertices with gauge fields with an inverse power of momentum. In this way, the 
presence of the Gribov horizon leads not only to the modification of the gluon 
propagator, but also to the presence of new diagrams.

The presence of mixed propagators lead to a more intricate relation between the connected 
and the one-particle irreducible (1PI) amplitudes. As discussed in the appendix of 
\cite{Mintz:2017qri}, the ghost-gluon vertex can be written in terms of the ghost and 
the gluon propagators, and also in terms of 1PI kernels that mix the Faddeev-Popov 
ghosts $\bar c$ and $c$, and Zwanziger's auxiliary fields $\bar\varphi$ and $\varphi$. 
Explicitly, 
\begin{widetext}
\begin{eqnarray}
 \vev{A_\mu^a(k)\,\bar c^b(p)\,c^c(q)} = G(p)G(q)D_{AA}(k)P_{\mu\nu}(k)
 \left\{\frac{\delta^3\Gamma}{\delta A_\nu^a(-k)\delta c^b(-p)\delta \bar c^c(-q)}
 +
 \frac{2g\gamma^2f^{ade}}{k^2+\mu^2}\frac{\delta^3\Gamma}{\delta c^b(-p)\delta \bar c^c(-q)\delta \varphi_\nu^{de}(-k)}\right\}_{q=-p-k}\nonumber\\
\end{eqnarray}
\end{widetext}
or, in a compact notation,
\begin{equation}\label{eq:Acc-local-shorthand-notation}
  \frac{\vev{A\,\bar c\, c}_c}{(\vev{\bar c\,c}_c)^2\vev{AA}_c} = \Gamma_{A\,\bar c\,c} + \frac{\vev{A\,\varphi}_c}{\vev{A\,A}_c}\Gamma_{\varphi\,\bar c\,c} 
  +\frac{\vev{A\,\bar\varphi}_c}{\vev{A\,A}_c}\Gamma_{\bar\varphi\,\bar c\,c}.
\end{equation}

Due to translational invariance, the three-point function depends on two independent 
momenta: $p$ (the antighost momentum) and $k$ (the gluon momentum). For generic momenta, 
the tensor structure of the ghost-gluon vertex is such that there are five independent
scalar functions \cite{Ball:1980ax}. An important feature of these expressions is that 
the RGZ expressions respect the so-called Taylor kinematics \cite{Taylor:1971ff}, 
\begin{eqnarray}\label{eq:Taylor-kin-antighost}
 (\Gamma_{A\,\bar c\,c})^{abc}_\mu(p,0,-p) = 0\,,
\end{eqnarray}
and the non-renormalization theorem of the ghost-gluon vertex, namely 
\begin{eqnarray}\label{eq:Taylor-kin-ghost}
 (\Gamma_{A\,\bar c\,c})^{abc}_\mu(-p,p,0) = -i gf^{abc}p_\mu,
\end{eqnarray}
which follow from the Ward identities of the theory.

In order to simplify our analysis, we 
shall restrict ourselves to the {\it soft-gluon limit}, that is, $k\rightarrow0$. As
a result, many simplifications take place (for more details on the calculation, see 
\cite{Mintz:2017qri}). First, the tensor structure of the vertex simplifies in such 
a way that only two of the Ball-Chiu functions survive. Second, the expressions of 
all diagrams simplify considerably, even though the remaining expression still 
demands some analytical work, mainly due to the form (\ref{eq:gluon-propagator}) 
of the RGZ gluon propagator. And third, diagram (IV) of Fig. 
\ref{fig:feyndiags} (which corresponds to the kernels $\Gamma_{\bar\varphi\bar c c}$
and $\Gamma_{\varphi\bar c c}$) vanishes in this limit. Therefore, in the soft-gluon
limit, one ends up having to calculate only one 1PI function ($\Gamma_{A\bar c c}$),
which receives contributions from all diagrams in Fig. 1, except IV (that vanishes).

At the end of the day, we find for $\Gamma_{A\bar c c}$ the following analytic expression
\begin{widetext}
\begin{eqnarray}\label{eq:resultGamma}
 [\Gamma_{A\bar c c}^{(1)}(0,p,-p)]^{abc}_{\mu} 
&=&ig^3\frac{Nf^{abc}}{2}\bigg\{R_+ J_\mu(a_+;p) + R_- J_\mu(a_-;p)
+2R_+^2K_\mu(a_+,a_+;p) + 2R_-^2K_\mu(a_-,a_-;p) + \nonumber\\
&&\left.+4R_+R_- K_\mu(a_+,a_-;p) +
\frac{N}{2}\left(\frac{g\gamma^2}{a_+^2-a_-^2}\right)^2
\left[K_\mu(a_+,a_+;p)+K_\mu(a_-,a_-;p)-\right.\right.\nonumber\\
&&-2K_\mu(a_+,a_-;p)\big]\bigg\}\,,
\end{eqnarray}
\end{widetext}
where the master integrals
\begin{equation}
 J_\mu(m_1;p):=\int_\ell\,
\frac{1}{\ell^2}\frac{1}{\ell^2+m_1^2}\frac{p^2\ell^2 - (p\cdot\ell)^2}{[(\ell-p)^2]^2}\,(\ell-p)_\mu,
\end{equation}
related to diagram (I), and
\begin{widetext}
\begin{eqnarray}
 K_\mu(m_1,m_2;p)&:=&\int_\ell \frac{1}{(\ell+p)^2}\frac{1}{\ell^2+m_1^2}\frac{1}{\ell^2+m_2^2}
\,\left[\frac{\ell^2p\cdot p - (p\cdot\ell)^2}{\ell^2}\right]\ell_\mu
\end{eqnarray}
\end{widetext}
have been calculated analytically in the appendix of \cite{Mintz:2017qri}. 
Note that both integrals are finite for $d=4$ and therefore the results 
at this level do not have any dependence on the renormalization scale. The quantities 
$R_\pm$ and $-a_\pm^2$ are, respectively, the residues and the poles of the tree-level 
gluon propagator, i.e.,
\begin{eqnarray}\label{eq:gluon-prop-partial-frac}
D_{AA}(p^2) &=& \frac{p^2+M^2}{(p^2+m^2)(p^2+M^2)+\Lambda^4} \nonumber\\
&\equiv& \frac{R_+}{p^2+a_+^2} + \frac{R_-}{p^2+a_-^2},
\end{eqnarray}
with
\begin{eqnarray}
 a_+^2 &=& \frac{m^2+M^2+\sqrt{(m^2-M^2)^2-4\Lambda^4}}{2},\nonumber\\
 a_-^2 &=& \frac{m^2+M^2-\sqrt{(m^2-M^2)^2-4\Lambda^4}}{2},\nonumber\\
 R_+ &=& \frac{m^2-M^2+\sqrt{(m^2-M^2)^2-4\Lambda^4}}{2\sqrt{(m^2-M^2)^2-4\Lambda^4}},\nonumber\\
 R_- &=& \frac{-m^2+M^2-\sqrt{(m^2-M^2)^2-4\Lambda^4}}{2\sqrt{(m^2-M^2)^2-4\Lambda^4}} \nonumber\\
 &=& 1-R_+,
\end{eqnarray}
where $\Lambda^4 = 2g^2N_c\gamma^4$. The other relevant propagators necessary to 
calculate the vertex function at the one-loop approximation are the mixed propagator
\begin{eqnarray}
\langle A_{\mu}^a(p)\varphi_{\nu}^{bc}(-p)\rangle
&=&\frac{g\gamma^2f^{abc}\;\;\mathcal{P}_{\mu\nu}(p) }{p^4+p^2(m^2+M^2)+m^2M^2+2Ng^2\gamma^4}\nonumber\\
&=& g\gamma^2f^{abc}\mathcal{P}_{\mu\nu}(p)\frac{D_{AA}(p)}{p^2+M^2}
\end{eqnarray}
and the ghost propagator
\begin{eqnarray}
 \vev{\bar c^a(p)\,c^b(-p)} =  \frac{\delta^{ab}}{p^2}.
\end{eqnarray}
The tree-level vertices used in the calculation are the three-gluon vertex
\begin{widetext}
\begin{eqnarray}
 ^{tree}[\Gamma_{AAA}(k,p,q)]^{abc}_{\mu\nu\rho}
 &=& igf^{abc}\left[(k_\nu-q_\nu)\delta_{\rho\mu} + (p_\rho-k_\rho)\delta_{\mu\nu} + (q_\mu-p_\mu)\delta_{\nu\rho} \right],
\end{eqnarray}
\end{widetext}
the ghost-gluon vertex,
\begin{eqnarray}
^{tree}[\Gamma_{A\bar c c}(k,p,q)]^{abc}_{\mu} &=& -igf^{abc}p_\mu
\end{eqnarray}
and the gluon-phi vertex
\begin{eqnarray}
 ^{tree}[\Gamma_{A\bar \varphi \varphi}(k,p,q)]^{abcde}_{\mu\nu\rho}
 &=& -igf^{abd}\delta^{ce}\delta_{\nu\rho}p_\mu.
\end{eqnarray}

In the next section, we will compare our results numerically with those obtained 
from other methods, namely lattice Monte Carlo simulations and Schwinger-Dyson 
equations.

\section{Numerical results}\label{sect:results}

In order to compare our results with other approaches, let us first fix our parameters.
The RGZ parameters $m^2$, $M^2$ and $\gamma^4$ can be fixed by two possible strategies.
The first one is a self-consistent method in which all three parameters are related to 
$\Lambda_{QCD}$ by calculating the RGZ effective potential to some approximation
(for example, up to one loop), as outlined in \cite{Vandersickel:2011zc}. In this 
work, we follow a second, more practical, approach, which is to directly fit the three 
RGZ parameters from lattice results on the gluon propagator. In
\cite{Oliveira:2012eh}, the gluon propagator has been fitted to the function
\begin{eqnarray}\label{eq:prop-fit}
 D_{fit}(p^2) = \frac{p^2 + a}{p^4 + bp^2 + c},
\end{eqnarray}
and the values for the parameters in Table \ref{table:parameters} have been found. 
Comparing expressions (\ref{eq:gluon-prop-partial-frac}) and (\ref{eq:prop-fit}) 
for the gluon propagator, the values for the RGZ parameters correspond to $M^2=4.473\,$
GeV$^2$, $m^2=-3.768\,$GeV$^2$, and $2g^3N_c\gamma^4=17.25\,$GeV$^4$, valid for a pure 
gauge $SU(3)$ theory.

\begin{table}[h!]
  \begin{center}
    \begin{tabular}{c|c|c}
       $a$ (${\rm GeV}^2$) \hspace{.2cm} & \hspace{.2cm} $b$ (${\rm GeV}^2$) \hspace{.2cm} & \hspace{.2cm} $c$ (${\rm GeV}^4$) \hspace{.2cm} \rule{0pt}{3ex} \\
\hline\hline
 \rule{0pt}{2ex} 
   4.473(21)  & 0.704(29)   & 0.3959(54)
\end{tabular}  
\end{center}
    \caption{RGZ parameters fitted from lattice results for SU(3) \cite{Oliveira:2012eh}, with error estimates in parenthesis.}
    \label{table:parameters}
\end{table}

\begin{figure}[h!]
\includegraphics[height=5.5cm]{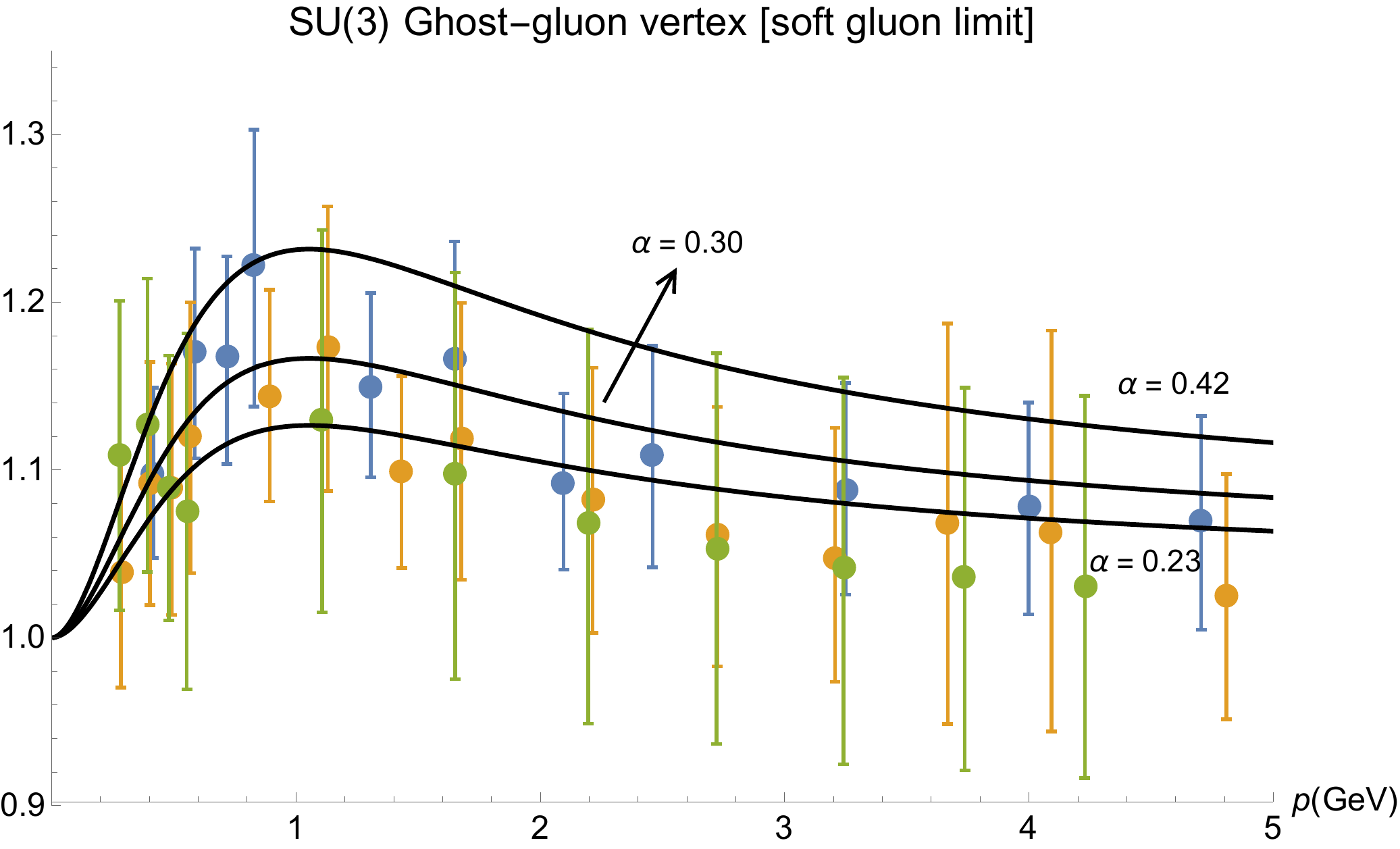}
\caption{The form factor $\Gamma_{A\bar c c}(p)$ of the SU(3) ghost-gluon vertex in the soft-gluon limit as a function of the antighost momentum for $d=4$ is compared to lattice simulations. The solid lines 
represent our results for different values of the strong coupling: $\alpha=\frac{g^2}{4\pi}=0.23,\, 0.3, \, 0.42$, from the bottom to the top curve, respectively. Data points correspond to lattice results from Ref. \cite{Ilgenfritz:2006he}.
}\label{fig:SU(3)compLattice}
\end{figure}

In Fig. \ref{fig:SU(3)compLattice}, we compare our result for the ghost-gluon 
vertex function $\Gamma_{A\bar c c}(p)$, defined at the r.h.s. of
\begin{eqnarray}
 [\Gamma_{A\bar c c}(0,p,-p)]^{abc}_{\mu} 
&=&-ig f^{abc} p_{\mu}\Gamma_{A\bar c c}(p),
\end{eqnarray}
at the soft-gluon limit with lattice estimates from \cite{Ilgenfritz:2006he}, 
for fixed coupling $\alpha=\frac{g^2}{4\pi}=0.23,\, 0.3,\, 0.42$. The solid lines 
represent the RGZ results. Notice that, in the large $p$ limit, the curves approach 
the standard perturbative result for a fixed coupling,
\begin{eqnarray}
 \Gamma_{A\bar c c}(p)= 1+ g^2N_c \frac{3}{64\pi^2},
\end{eqnarray}
while they go to the perturbative limit at low momenta, passing through a maximum 
at a momentum scale of the order of 1 GeV.

\begin{figure}
\includegraphics[height=5.5cm]{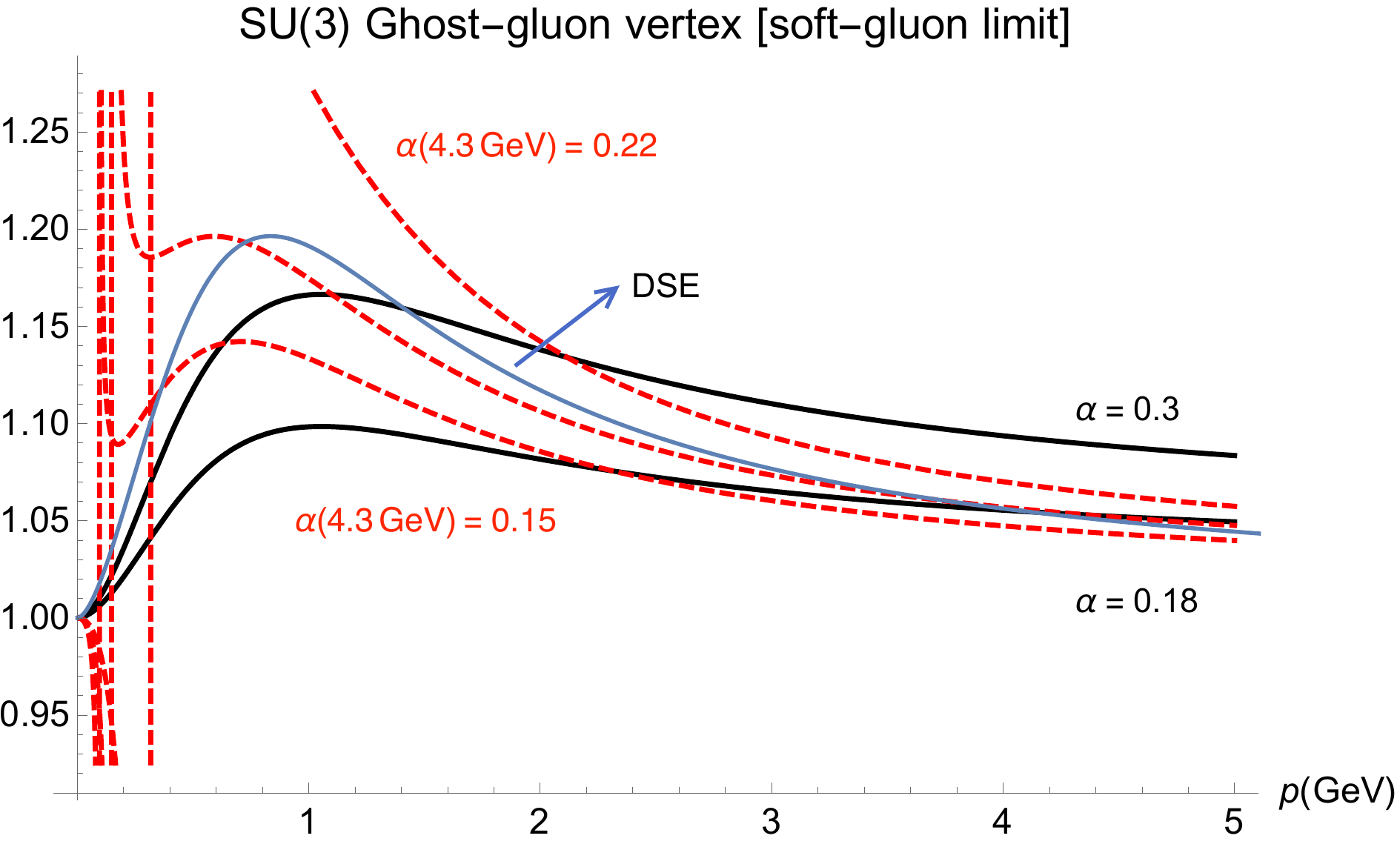}
\caption{The scalar function $\Gamma_{A\bar c c}(p)$ of the SU(3) ghost-gluon vertex in the soft-gluon limit as a function of momentum for $d=4$.
Solid thick (black) lines represent our results for different values of the strong coupling: $\alpha=\frac{g^2}{4\pi}=0.18$ (bottom) and $\alpha=0.3$ (top), while the dashed (red) lines include
a one-loop perturbative running coupling with different renormalization conditions: $\alpha(\mu=4.3 {\rm GeV})=0.15,\,  0.18, \, 0.22$, from the bottom to the top curve, respectively. 
A DSE result from Ref. \cite{Aguilar:2013xqa} is represented as the thin (blue) line. }\label{fig:SU(3)compDSErunning}
\end{figure}

We also consider the running coupling constant
\begin{eqnarray}\label{Eq:Running}
 g^2(p) = \frac{g^2(\mu)}{1+\frac{11N_c}{3}\,\frac{g^2(\mu)}{8\pi^2}\log(\frac{p}{\mu})}\,,
\end{eqnarray}
which corresponds to the standard one-loop YM beta function in the Modified 
Minimal Subtraction scheme, cf. \cite{Peskin:1995ev}, and compare to our results of 
fixed coupling and to the Schwinger-Dyson approach \cite{Aguilar:2013xqa}. We see 
that our result (dashed line in Figure \ref{fig:SU(3)compDSErunning}) is 
qualitatively comparable to the DSE vertex, with the running coupling somewhat 
interpolating between the fixed coupling values $\alpha=0.18$ and $\alpha=0.3$. 
For lower momenta, the perturbative running coupling diverges due to the Landau 
pole, making the vertex function also divergent. We believe that an adequate 
IR renormalization scheme, such as the one used in \cite{Pelaez:2013cpa}, could 
lead to a mass-dependent beta function in RGZ, and thus a possible absence of 
the Landau pole, leading to a smooth vertex function in a full treatment including 
the renormalization group improvement.

\section{Summary and some final remarks}\label{sect:remarks}

In covariantly quantized gauge theories, one has to deal with the problem of 
gauge copies (i.e., the Gribov problem \cite{Gribov:1977wm,Vandersickel:2012tz}) 
as the infrared regime is approached. The resulting GZ action is unstable 
with respect to the formation of dimension two condensates, which effectively 
lead to the RGZ effective theory, which presents a massive gluon behavior.
The gluon propagator at the RGZ effective theory displays a nice agreement 
with lattice gluon propagator already at tree level. This can be interpreted 
as the RGZ theory providing a sort of nontrivial vacuum over which the gauge 
fields evolve. One can calculate further corrections over this ``vacuum'' by 
considering perturbative correction to correlation fuctions, por example.
Using parameters for the tree-level RGZ propagator fitted from lattice YM results
(and only that!), we calculated the ghost-gluon vertex at the soft gluon
limit ($k\rightarrow0$) and found qualitative agreement with other methods, 
such as the lattice itself and Dyson-Schwinger equations.

Next, we intend to explore the behavior of the ghost-gluon vertex in the RGZ 
framework for any momentum configuration. Besides, other YM vertices like 
the three- and four-gluon vertices would also be very interesting to study.
The inclusion of matter (quark) fields and the exploration of the gauge
dependence of the vertex are also possible extensions of this work.

\section*{Acknowledgements}

B.~W.~M.~ would like to thank the organizers of XIV Hadron Physics for a 
very pleasant and well organized conference.  
This work has been partially supported by CNPq, CAPES, and 
FAPERJ. It is a part of the project INCT-FNA Proc. 
464898/2014-5. A.D.P acknowledges funding by the DFG, Grant Ei/1037-1.

\bibliography{RGZ-bibliography}

\end{document}